\documentclass[singlecolumn,epjc3]{svjour3}
\smartqed  
\RequirePackage{graphicx}
\journalname{Eur. Phys. J. A}

\begin{document}

\title{Modelling of anisotropic compact stars of embedding class one}

\author{Piyali Bhar\thanksref{e1,addr1}
\and
S.K. Maurya\thanksref{e2,addr2}
\and
Y. K. Gupta\thanksref{e3,addr3}
\and
Tuhina Manna\thanksref{e4,addr4}.}

\thankstext{e1}{e-mail:piyalibhar90@gmail.com }
\thankstext{e2}{e-mail: sunil@unizwa.edu.om}
\thankstext{e3}{e-mail: kumar001947@gmail.com}
\thankstext{e4}{e-mail: tuhinamanna03@gmail.com}

\institute{Department of
Mathematics,Government General Degree College, Singur, Hooghly 712 409, West Bengal,
India\label{addr1}
\and
Department of Mathematical \& Physical Sciences,
College of Arts \& Science, University of Nizwa, Nizwa, Sultanate
of Oman\label{addr2}
\and
Department of Mathematics, Raj Kumar Goel Institute of Technology, Ghaziabad, U.P.(India)\label{addr3}
\and
Department of
Commerce (Evening) , St. Xavier's College, 30, Mother Teresa
Sarani,  Kolkata 700 016, West Bengal, India\label{addr4}}

\date{Received: date / Accepted: date}

\maketitle

\begin{abstract}
In the present article, we have constructed static anisotropic compact star models of Einstein field equations for the spherical symmetric metric of embedding class one. By assuming the particular form of metric function $\nu$, We have solved the Einstein field equations for anisotropic matter distribution. The anisotropic models are representing the realistic compact objects such as SAX J 1808.4-3658 (SS1), Her X - 1, Vela X-12, PSR J1614-2230 and Cen X - 3. We have reported our results in details for compact star Her X-1 on the ground of physical properties such as pressure, density, velocity of sound, energy conditions, TOV equation and red-shift etc. Along with these, we have also discussed about stability of the compact star models. Finally we made the comparison between our anisotropic stars with the realistic objects on the key aspects as central density, central pressure, compactness and surface red-shift.
\end{abstract}


\maketitle

\section{Introduction}

Compact objects such as white dwarf stars, neutron stars and
black holes represent the final stage of a star's evolution.
Stars are born in gaseous nebulae in which clouds of hydrogen
coalesce, becomes highly compressed and heated through the
gravitational interaction and at a temperature of about $10^7$ K,
a nuclear reaction begins converting hydrogen into the next
heavier element, helium, and eventually into more heavier
elements, thus releasing a large quantity of electromagnetic energy
(light). However, as the star burns hotter and ignites heavier
elements which accumulate in the core, electromagnetic pressure
becomes less and less effective against gravitational collapse
till the star collapses under its own weight resulting in
creation of these stellar remnants. Recent advances in
observation cosmology have explored interesting facts regarding
evolution of these compact objects and revealed many of their
features from the study of their emission spectra and rotational
frequencies. But it is still a challenge to measure some of the
other important parameters like mass, internal composition,
radii, etc. which cannot be inferred from direct observational
data. This is when the theory of general relativity comes into
play. Theoretical relativistic stellar models have been known to
successfully predict many crucial properties of compact objects
which were impossible to determine otherwise.\par
In our model we assume the interior of a compact star to be an
anisotropic fluid since high densities of the order $10^{15}
g.cm^{-3}$ causes nuclear fluids to be anisotropic in nature \cite{Ruderman}. In case of anisotropy the radial and tangential
pressures are not equal ($p_r \neq p_t$). Note that the
anisotropy factor $\Delta=p_t-p_r$ increases rapidly with
increase in radial distance but vanishes at the centre of our
stellar model. Anisotropy in nature may occur due to presence of
a solid core or type 3A fluid or type P superfluid \cite{Kippenhahn and Weigert} or may result from
different kind of phase transition, rotation, magnetic stress,
pion condensation, etc. Here we would like to mention that Mak and
Harko\cite{Mak and Harko},and Sharma and Maharaj\cite{Sharma and
Maharaj} suggest that anisotropy is a sufficient condition in the
study of dense nuclear matter with strange star. In an earlier
work Ponce de Leon\cite{Ponce` de Leon} obtained two new exact
analytical solutions to Einstein's field equations for static
fluid sphere with anisotropic pressures. In addition to that,
Herrera and Santos \cite{Herrera and Santos(1997a)}
provided an exhaustive review on the subject of anisotropic
fluids. Komathiraj \& Maharaj \cite{maha1} presented exact solutions to the Einstein-Maxwell system of equations in
spherically symmetric gravitational fields with a specified form of the electric
field intensity. Considering Vaidya-Tikekar metric,  Chattopadhyay {\em et al.}\cite{cha} obtained a class of solutions of the Einstein-Maxwell equations for a charged static fluid sphere. Realistic models of relativistic radiating stars undergoing gravitational collapse which have vanishing Weyl tensor components are proposed by  Misthry {\em et al.} \cite{mis}.
 The possibility of forming of anisotropic compact stars from the cosmological constant as one of the competent candidates of dark energy Hossein \cite{hos}
 obtained a new model of compact star by using Krori and Barua metric. Bhar {\em et al.} \cite{pb} studied the behavior of static spherically symmetric relativistic objects with locally anisotropic matter distribution considering the Tolman VII form for the gravitational potential $g_{rr}$ in curvature coordinates together with the linear relation between the energy density and the radial pressure. Bhar \cite{pb2}  proposed a new model of an anisotropic strange star which admits the Chaplygin equation of state. The model is developed by assuming the Finch–Skea ansatz. Bhar \& Rahaman \cite{bhar} proposed a new model of dark energy star consisting of five zones, namely, the solid core of constant energy density, the thin shell between core and interior, an inhomogeneous interior region with anisotropic pressures, a thin shell, and the exterior vacuum region. Various physical properties have been discussed.
 B\"{o}hmer and Harko \cite{harko1}derived upper and lower limits for the basic physical parameters $viz.$ mass-radius ratio, anisotropy,
redshift and total energy for arbitrary anisotropic general relativistic matter distributions in the
presence of cosmological constant. They have shown that anisotropic compact stellar type objects can be much more
compact than the isotropic ones, and their radii may be close to their corresponding Schwarzschild
radii. In this connection some other useful solution for compact star in different context have been obtained by several authors which can be seen in following references:{Mafa Takisa et al.\cite{mafa2014a,mafa2014b}, Ngubelanga et al.\cite{ngubelanga2015}, Sunzu\cite{sunzu2014}, Malaver\cite{malaver2014,malaver2016}, Pant and Maurya \cite{pant2012}, Maurya et al.\cite{mauryagupta2012,mauryagupta2015a,mauryagupta2015b}. \\
In present paper we have investigated a metric of embedding class one.The idea about class of metric is that if  we embedded our space-time into higher dimensional flat space time then this extra dimension is called the class of the metric \cite{Eddington1924}.  This idea is attracting much attention again due to proposal of Randall and Sundrum \cite{Randall} and Anchordoqui and Berglia \cite{Anchordoqui} discussions. The physical meaning of higher dimensional space cannot be provided by the general theory of relativity. But it gives the new characterizations of gravitational fields, which can relate to physics. The group of motions of flat embedding space have linked the internal symmetries of elementary particle physics \cite{Rayski}. However the higher dimensional space time are utilized for studying about the singularity of space time. In recent days, Pavsic and Tapia \cite{Pavsic} have provided the reference about the application of embedding into general relativity, extrinsic gravity, strings and membranes and new brane world.   To study the evolution problem for Einstein's equations using by embedding diagrams of Schwarzschild space and Misner's wormhole manifold have been discussed by Treibergs \cite{Treibergs}.
 It is well known that every $n$
dimensional Riemannian manifold $V_n$ can be isometrically
embedded into some pseudo-Euclidean space of $m$ dimensions where
$m= \frac{n(n+1)}{2} $. The embedding class of $V_n$ is the
minimum number of extra dimensions required by the
pseudo-Euclidean space, which is obviously equal to $\it{m-n =\frac{
n(n-1)}{2} }$ . For the $4$ dimensional Minkowski spacetime, the
embedding class is obviously $6$.
Another feature of class one
metric is that the metric functions $\lambda$ and $\nu$ are
dependent on each other. We would like to mention a very recent
work by Maurya et. al.\cite{Maurya(2015a),Maurya(2016)} on anisotropic compact
stars of embedding class one. However Maurya et al. \cite{Maurya(2015b),Maurya(2016a)} and Singh et al.\cite{ntn1,ntn2,ntn3} have given the methodology for constructing the anisotropy with the help of metric functions.\par
 In this paper we have calculated relevant values of parameters for compact star such as Her
X-1, SAX J 1808.4-3658(SS1), Vela X-12, PSR J1614-2230 and Cen
X-3  and obtained some agreeable results.
We have also analyzed all physical features in details and
provided sample figures to support our data. The mass, redshift
and compactness factor are within optimal ranges while the model
is stable under the action of hydrostatic, anisotropic and
gravitational forces. The subliminal velocity of sound and the
adiabatic index reconfirms the stability of our model.\par
This paper is divided in the following sections; in section $2$ the basic
Einstein's field equations are solved, in section $3$ the values
of constants $B$, $C$ and $F$ are obtained from the matching condition, next in section $4$ a
physical analysis of the model is done. The mass-radius relation
is explored in section $5$ including the redshift and compactness.
In the following section $6$ we have studied the energy
conditions. Section $7$ is devoted to the study of Stability of
the model and some discussions are made in the final section.

\section{Basic field equations and Anisotropic solutions}
\subsection{Basic field equations:}
To describe the interior of a static and spherically symmetry object the line element can be taken in the Schwarzschild co-ordinate $(x^{a})=(t,r,\theta,\phi)$ as,
\begin{equation}
ds^{2}=e^{\nu(r)}dt^{2}-e^{\lambda(r)}dr^{2}-r^{2}\left(d\theta^{2}+\sin^{2}\theta d\phi^{2} \right) \label{1}
\end{equation}
Where $\nu$ and $\lambda$ are functions of the radial coordinate `r' only.\\
Now the spacetime (\ref{1}) represent the spacetime of class one, if the spacetime (\ref{1}) satisfies the Karmarkar condition \cite{kar} as:
\begin{equation}
R_{1414}=\frac{R_{1212}R_{3434}+ R_{1224}R_{1334}}{R_{2323}}
\end{equation}
with $R_{2323}\neq 0$ \cite{pandey}. \\
Now the above components of $R_{hijk}$ for metric (1) are:
\begin{eqnarray}
R_{2323}=\,r^2\,sin^{2}\theta \,\left[1-e^{-\lambda}\right] \nonumber
\end{eqnarray}

\begin{eqnarray}
R_{1212}=\frac{1}{2}\,\lambda'\,r \nonumber
\end{eqnarray}

\begin{eqnarray}
R_{1224}=0   \,\,\,~\,~\,\nonumber
\end{eqnarray}

\begin{eqnarray}
R_{1414}=e^{\nu}\left[\frac{1}{2}\,\nu''+\frac{1}{4}\,{\nu'}^{2}-\frac{1}{4}\,\lambda'\,\nu'\right] \nonumber
\end{eqnarray}

\begin{eqnarray}
R_{2424}=\frac{1}{4}\,\nu'\,\beta' \,e^{\nu-\lambda} \nonumber
\end{eqnarray}

\begin{eqnarray}
R_{3434}=sin^{2}\theta\,R_{2424} \nonumber
\end{eqnarray}

On inserting the values above components in Eq.(2), we get the following following differential equation:

\begin{equation}
\frac{\lambda'\nu'}{1-e^{\lambda}}=-2(\nu''+\nu'^{2})+\nu'^{2}+\lambda'\nu'.
\end{equation}
with $e^{\lambda}\neq 1.$ Solving equation (3)we get,
\begin{equation}
e^{\lambda}=1+F\nu'^{2}e^{\nu}
\end{equation}
where $F\neq0$ is an arbitrary constant.\\
The Eq.(4) represents the class $1$ condition for metric (1).
We assume that the matter within the star is anisotropic in nature and correspondingly the energy-momentum tensor is described by,
\begin{equation}
T_{\nu}^{\mu}=(\rho+p_r)u^{\mu}u_{\nu}-p_t g_{\nu}^{\mu}+(p_r-p_t)\eta^{\mu}\eta_{\nu}
\end{equation}
with $ u^{i}u_{j} =-\eta^{i}\eta_j = 1 $ and $u^{i}\eta_j= 0$, the vector $u_i$ being the fluid 4-velocity and $\eta^{i}$ is the spacelike vector which is orthogonal to $ u^{i}$. Here $\rho$ is the matter density, $p_r$ is the the radial and $p_t$ is transverse pressure of the fluid in the orthogonal direction to $p_r$.\\
Now for the line element (1) and the matter distribution (5) Einstein Field equations (assuming $G=c=1$) is given by,
\begin{equation}
	\frac{1-e^{-\lambda}}{r^{2}}+\frac{e^{-\lambda}\lambda'}{r}=8\pi\rho,	
\end{equation}
\begin{equation}
	\frac{e^{-\lambda}-1}{r^{2}}+\frac{e^{-\lambda}\nu'}{r}=8\pi p_{r},
\end{equation}
\begin{equation}
	e^{-\lambda}\left(\frac{\nu''}{2}+\frac{\nu'^{2}}{4}-\frac{\nu'\lambda'}{4}+\frac{\nu'-\lambda'}{2r} \right)=8\pi p_t,
\end{equation}

\subsection{Anisotropic solution for compact stars:}
Now we have to solve the Einstein field equations (6)-(8)with the help of equation (4). One can notice that we have four equation with $5$ unknowns namely $\lambda,\nu,\rho,p_r$ and $p_t$.\\
To solve the above set of equations let us take the metric co-efficient $g_{tt}$ proposed by Tolman \cite{tolman} as,
\begin{equation}
e^{\nu}=B(1+Cr^{2})^{4}
\end{equation}
Where $B$ and $C$ are positive constants.\\
As Lake \cite{lake2003} has suggested that for any physically acceptable model, the metric function $e^{\nu}$ should be monotonically increasing with increase of $r$ and it should be positive, free from singularity at centre with $\nu'(0)=0$. It is observed from Eq.(9) that our metric function $e^{\nu}$ is monotonic increasing function of $r$ and $e^{\nu(0)}=B$, $\nu'(0)=0$. This implies that $e^{\nu}$, given by equation (9), is physically acceptable.

Solving equation (4) and (9) we obtain,
\begin{equation}
e^{\lambda}=1 + 64 B C^{2}F r^{2} (1 + C r^{2})^{2}
\end{equation}
Now employing the values of $e^{\nu}$ and $e^{\lambda}$ to the Einstein's field equations $(6)-(8)$ we obtain the expression for matter density, radial and transverse pressure as,
\begin{equation}
8\pi\rho=\frac{64 B C^{2} F (1 + C r^{2}) \left[3 + C r^{2} \left\{7 + 64 B C F (1 + C r^{2})^{3}\right\}\right]}{[1 + 64 B C^2 F r^2 (1 + C r^2)^2]^2}
\end{equation}
\begin{equation}
8\pi p_r=\frac{8 C [1 - 8 B C F (1 + C r^2)^3]}{(1 + C r^2) [1 + 64 B C^2 F r^2 (1 + C r^2)^2]}
\end{equation}
\begin{equation}
8\pi p_t=\frac{8 C [1 + 2 C r^2 - 8 B C F (1 - C r^2) (1 + C r^2)^3]}{(1 + C r^2)^2 [1 + 64 B C^2 F r^2 (1 + C r^2)^2]^2}
\end{equation}
and the anisotropic factor $\Delta$ is obtained by using the pressure isotropy condition as,
\begin{equation}
8\pi \Delta=\frac{8 C^2 r^2 [1 - 48 B C F (1 + C r^2)^3 + 512 B^2 C^2 F^2 (1 + C r^2)^6]}{(1 + C r^2)^2 [1 + 64 B C^2 F r^2 (1 + C r^2)^2]^2}
\end{equation}

In this connection we want to mention that all static anisotropic solutions to Einstein equations in the spherically symmetric case can be obtained by a general method introduced by Herrera and co-workers \cite{he}. As shown in this reference, all such solutions are  produced by two generating functions. Accordingly,  the solution presented by the authors is a particular case of the general solution described in the reference above. For our model the expression for these two generating functions are given by,
\begin{eqnarray}
  y(x) &=& e^{\nu}=B(1+Cr^{2})^{4} \\
  \Pi &=& 8\pi \Delta=\frac{8 C^2 r^2 [1 - 48 B C F (1 + C r^2)^3 + 512 B^2 C^2 F^2 (1 + C r^2)^6]}{(1 + C r^2)^2 [1 + 64 B C^2 F r^2 (1 + C r^2)^2]^2}
\end{eqnarray}
\section{The values of constants $B, C$ and $F$}
To fix the values of the constants $B, C$ and $F$ we match our interior spacetime to the exterior schwarzschild line element given by
\begin{equation}
ds^{2}=-\left(1-\frac{2m}{r}\right)dt^{2}+\left(1-\frac{2m}{r}\right)^{-1}dr^{2}+r^{2}(d\theta^{2}+\sin^{2}\theta d\phi^{2})
\end{equation}
outside the event horizon $r>2m$, $m$ being the mass of the black hole.\\
using the continuity of the metric coefficient $e^{\nu},e^{\lambda}$ and $\frac{\partial g_{tt}}{\partial r}$ across the boundary we get the following three equations
\begin{equation}
1-\frac{2m}{r_b}=B(1+C r_b^{2})^{4}
\end{equation}
\begin{equation}
1+64BC^{2}Fr_b^{2}(1+Cr_b^{2})^{2}=\left(1-\frac{2m}{r_b}\right)^{-1}
\end{equation}
\begin{equation}
8BC(1+Cr_b^{2})^{3}=\frac{2m}{r_b^{3}}
\end{equation}

Solving the equations $(18)-(20)$ we obtain the values of $B,C$ and $E$ in terms of mass of radius of the compact star as,
\begin{equation}
B=\frac{\left(1-\frac{9m}{4r_b}\right)^{4}}{\left(1-\frac{2m}{r_b}\right)^{3}}
\end{equation}
\begin{equation}
C=\frac{1}{4}\frac{m}{r_b^{3}}\left(1-\frac{9m}{4r_b}\right)^{-1}
\end{equation}
\begin{equation}
F=\frac{1}{2}\left(\frac{m}{r_b^{3}}\right)^{-1}
\end{equation}

\section{Physical analysis of our present model}
To be a physically acceptable solution our present model should satisfy the following conditions:\\

1. The metric coefficient should be regular inside the stellar interior. From our solution we can easily check that $e^{\lambda(r=0)}=1$ and $e^{\nu(r=0)}=B$, a positive constant. To see the characteristic of the metric potential we plotted the graph of $e^{\nu}$ and $e^{\lambda}$ in fig.\ref{metric}. The profiles show that metric coefficients are regular and monotonic increasing function of $r$ inside the stellar interior.
\begin{figure}[h!]
    \centering
        \includegraphics[scale=.7]{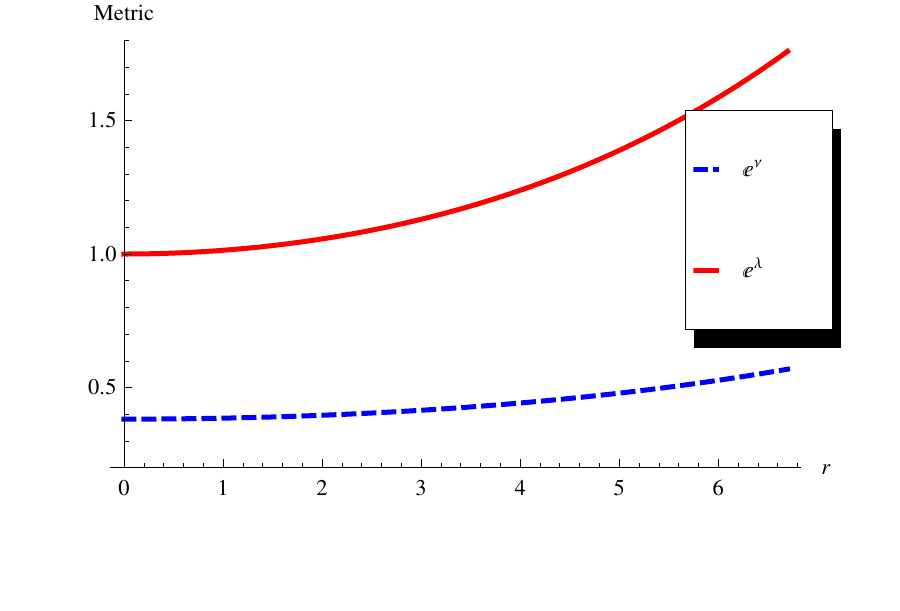}
       \caption{The metric potentials are plotted against r for the compact star Her X-1 by taking $B=0.382$, $C=2.34\times{10^{-3}} $ and $F=104.03$. The metric function $e^{\nu}$ is shown by dashed line (Blue color)and $e^{\lambda}$ is shown by solid line (red color).}\label{metric}
\end{figure}

2. The matter density, radial and transverse pressure should be positive inside the stellar interior for a physically acceptable model. The radial pressure should be vanish at the boundary of the star.\\

Moreover the central density and central pressure can be obtained as,
\[\rho_c=\rho(r=0)=\frac{24BC^{2}F}{\pi}\]
\[p_r(r=0)=p_t(r=0)=\frac{C(1-8BCF)}{\pi}\]
Therefore our model is free from central singularity. The behavior is shown in fig. 2.

\begin{figure}[h!]
    \centering
        \includegraphics[scale=.58]{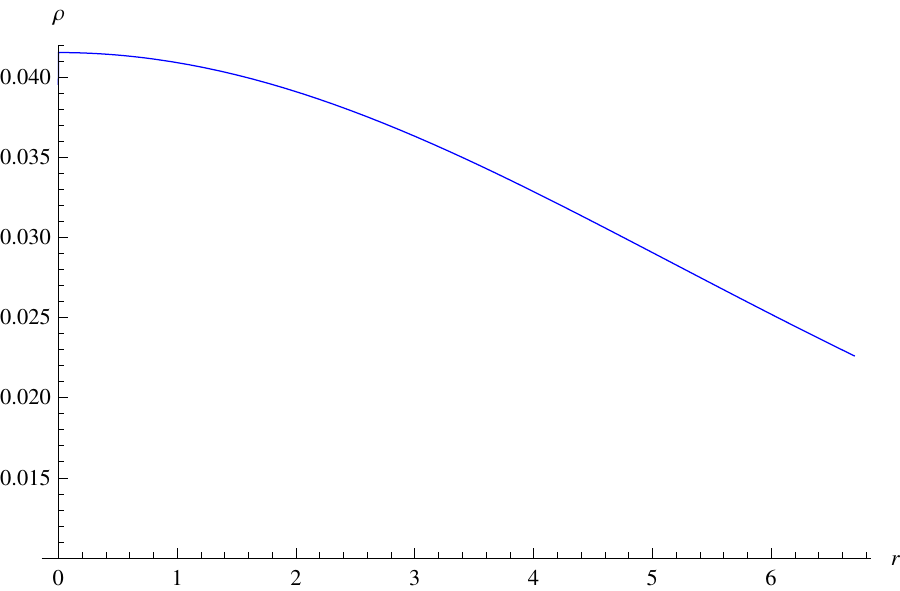} \includegraphics[scale=.65]{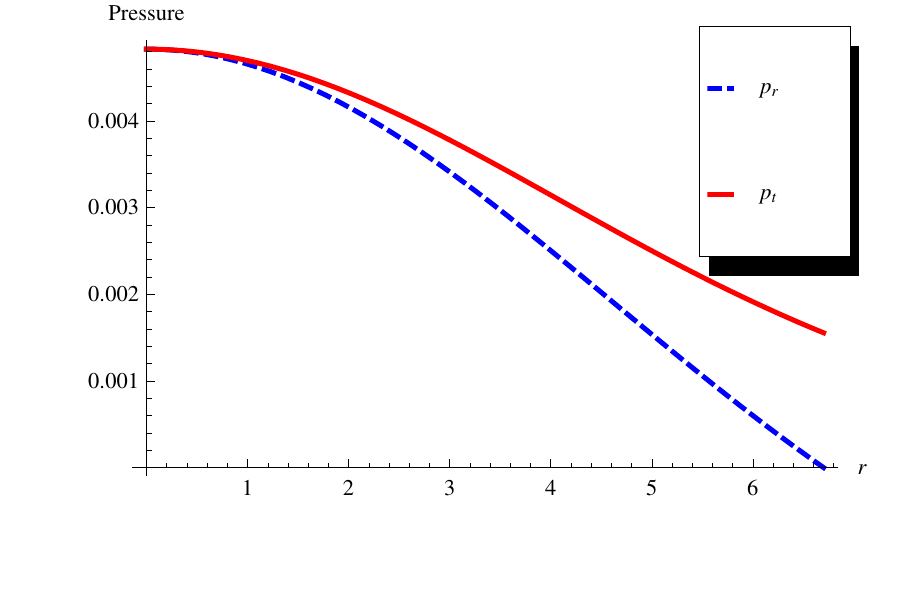}
       \caption{Variation of matter density (left panel) and pressures (right panel) is plotted against r for the compact star Her X-1 by taking the same values of the constant mentioned in fig. 1}\label{rho}
\end{figure}

Now for our model the gradient of matter density and radial pressure are obtained as,

\begin{eqnarray}
\frac{d\rho}{dr}=-\frac{256 B C^3 F r}{8\pi[1 + 64 B C^2 F r^2 (1 + C r^2)^2]^3}f_1(r)\\
\frac{dp_r}{dr}=-\frac{16 C^2 r}{8\pi(1 + C r^2)^2 [1 + 64 B C^2 F r^2 (1 + C r^2)^2]^2}f_2(r)
\end{eqnarray}
where,
\begin{eqnarray*}
f_1(r)&=&2048 B^2 C^3 F^2 r^2 (1 + C r^2)^6 +
   32 B C F (1 + C r^2)^2 (5 + 16 C r^2 + 23 C^2 r^4)\\
   &&-(5 +7 C r^2)\\
f_2(r)&=&1 - 512 B^2 C^2 F^2 (1 + C r^2)^6 +
   16 B C F (1 + C r^2)^2 (5 + 17 C r^2)
\end{eqnarray*}
\begin{figure}[h!]
    \centering
        \includegraphics[scale=.8]{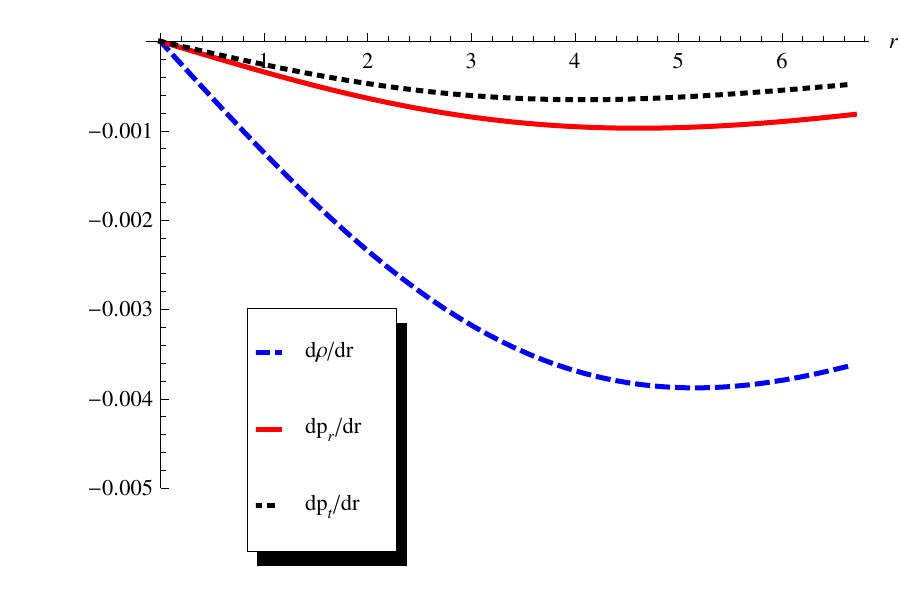}
       \caption{$\frac{dp_r}{dr}$(dotted line in black), $\frac{d\rho}{dr}$ dashed line in blue and $\frac{dp_t}{dr}$ (solid line in red)plotted against r for the compact star Her X-1 by taking the same values of the constant mentioned in fig. 1}\label{dp}
\end{figure}

We note that at the point $r=0$ both $\frac{d\rho}{dr}=0$ and $\frac{dp_r}{dr}=0$ and,
\[\frac{d^{2}\rho}{dr^{2}}=\frac{256 B C^3 F (5 - 160 B C F)}{8\pi}\]
\[\frac{d^{2}p_r}{dr^{2}}=-\frac{16 C^2 (1 + 80 B C F - 512 B^2 C^2 F^2)}{8\pi}\]

The profile of matter density, radial and transverse pressure are shown in fig. \ref{rho} respectively. The figures show that the matter density $\rho$, radial pressure $(p_r)$ and transverse pressure $(p_t)$ are monotonic decreasing function of $r$. All are positive for $0<r\leq r_b$ ($r_b$ being the boundary of the star) and both $\rho$ and $p_t$ are positive at the boundary where as the radial pressure vanishes there. The profile of $\frac{d\rho}{dr}$, $\frac{dp_r}{dr}$ and $\frac{dp_t}{dr}$ are plotted in fig.\ref{dp}. The plots show that both $\frac{d\rho}{dr}$, $\frac{dp_r}{dr}$ and $\frac{dp_t}{dr}$ are negative which once again verify that $\rho$, $p_r$ and $p_t$ are monotonic decreasing function of $r$. \\

3. The profile of the anisotropic factor $\Delta$ is shown against $r$ in fig.\ref{delta}. The anisotropic factor is positive and monotonic increasing function of $r$, which implies $p_t>p_r$, i.e., the anisotropic force is repulsive in nature and it helps to construct more compact object proposed by Gokhroo \& Mehra\cite{gokhroo}. Moreover at the center of the star the anisotropic factor vanishes which is also a required condition.\\

\begin{figure}[h!]
    \centering
        \includegraphics[scale=.8]{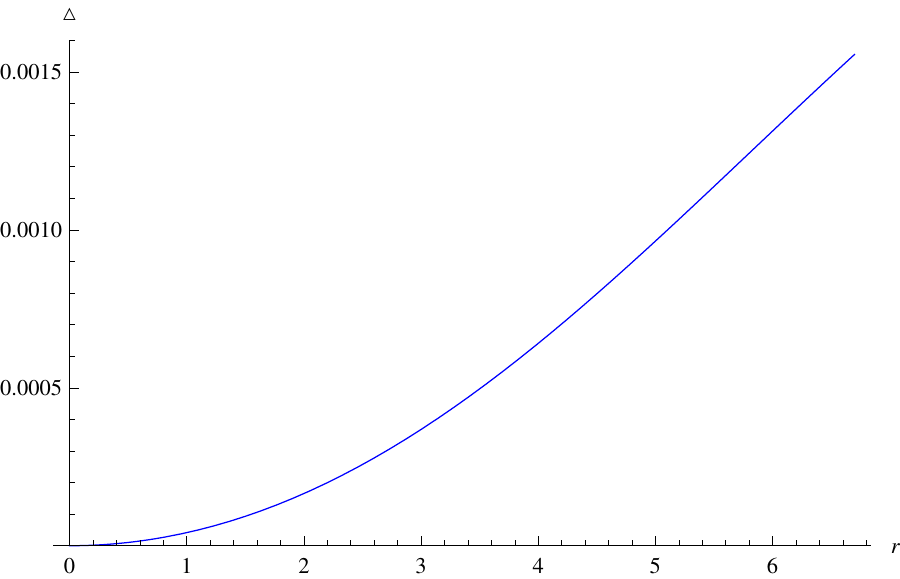}
       \caption{Variation anisotropic factor is plotted against r for the compact star Her X-1 by taking the same values of the constant mentioned in fig. 1}\label{delta}
\end{figure}

4. The parameters $\omega_r=p_r/\rho$ and $\omega_t=p_t/\rho$ are plotted in fig. \ref{omega}. From the figures it is clear that both $\omega_r$ and $\omega_t$ lie in the range $0<\omega_r,\omega_t<1$ implying that the underlying fluid distribution is non-exotic in nature \cite{saibal}

\begin{figure}[h!]
    \centering
        \includegraphics[scale=.8]{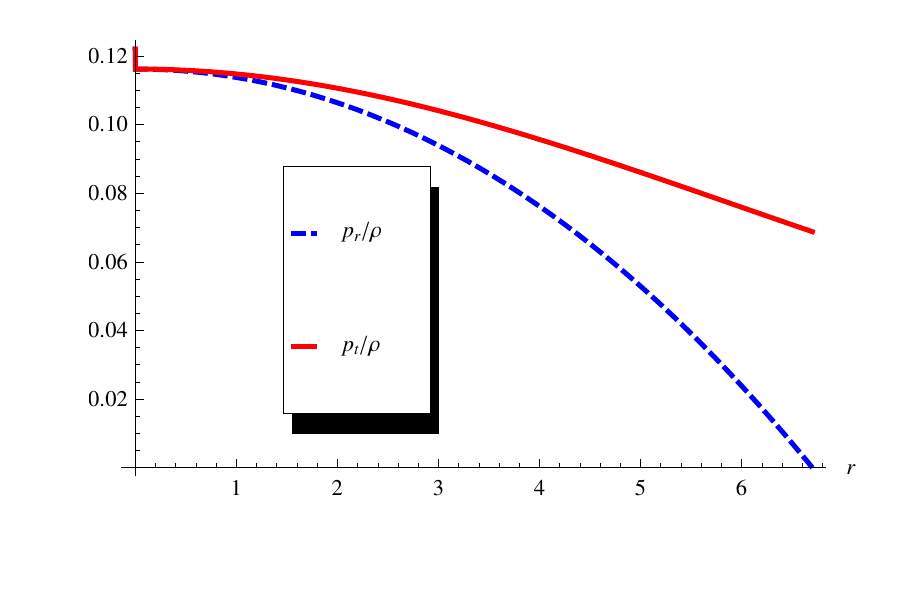}
       \caption{$p_r/\rho$ (dashed line in blue) and $p_t/\rho$ (solid line in red) plotted against r for the compact star Her X-1 by taking the same values of the constant mentioned in fig. 1}\label{omega}
\end{figure}

\section{Mass-radius relation}
The mass of the compact star is obtained as,
\begin{equation}
m(r)=\int_0^{r}4\pi\rho r^{2}dr=\frac{32 B C^2 F r^3 (1 + C r^2)^2}{1 + 64 B C^2 F r^2 (1 + C r^2)^2}
\end{equation}
The profile of the mass function is plotted against r in fig \ref{mass}. The profile shows that mass is an increasing function of r and $m(r)>0$ for $0<r<r_b$, so it is regular everywhere inside the stellar interior moreover at the center of the star the mass function vanishes.\\
The compactification factor of the star is obtained from the formula
\begin{equation}
u(r)=\frac{m(r)}{r}=\frac{32 B C^2 F r^2 (1 + C r^2)^2}{1 + 64 B C^2 F r^2 (1 + C r^2)^2}
\end{equation}
\begin{figure}[htbp]
    \centering
        \includegraphics[scale=.7]{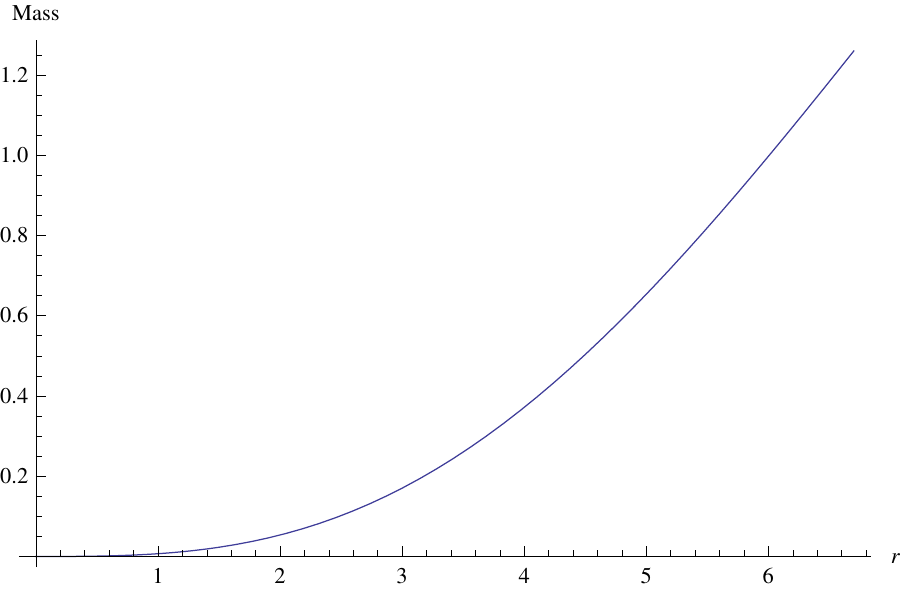}
       \caption{The mass function is plotted against r for the compact star Her X-1 by taking the same values of the constant mentioned in fig. 1}\label{mass}
\end{figure}

\begin{figure}[htbp]
    \centering
        \includegraphics[scale=.7]{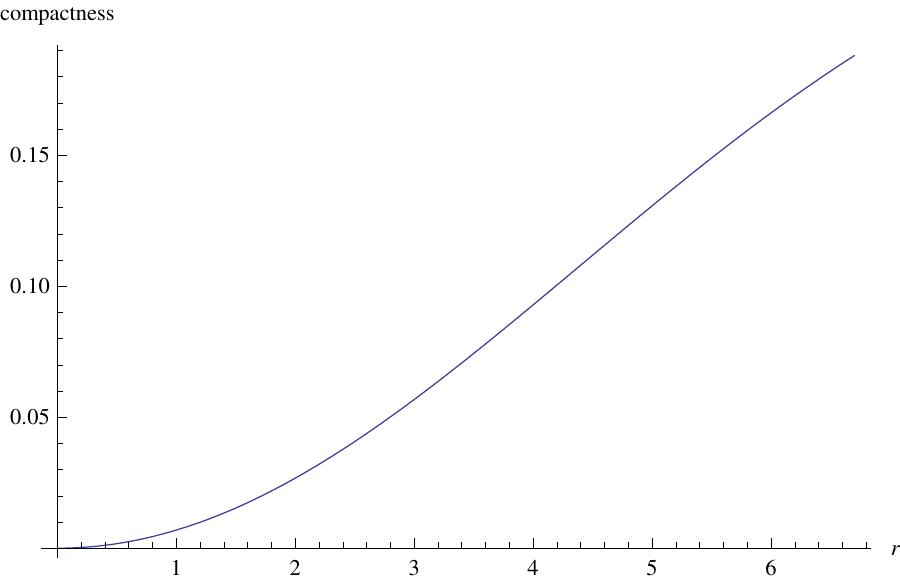}
       \caption{The compactness factor is plotted against r for the compact star Her X-1 by taking the same values of the constant mentioned in fig. 1}\label{ur}
\end{figure}

The profile of $u(r)$ is plotted against r in fig. \ref{ur}. The figure shows that $u(r)$ is also a monotonic increasing function of $r$.
Buchdahl \cite{buch} proposed that for a compact star twice the ratio of mass to the radius should be $<\frac{8}{9}.$ The compactness factor of some compact stars are calculated from our model which is shown in Table 1. The table shows that our model satisfy the Buchdahl's condition.\\
The surface redshift of a compact star $z_s$ is obtained from the following formula.
\begin{equation}
z_s=\left(1-2u\right)^{-\frac{1}{2}}-1
\end{equation}
For our model of compact star it becomes,
\[z_s=\sqrt{1 + 64 B C^2 F r^2 (1 + C r^2)^2}-1\]
We have obtained the value of surface redshift for different compact star which are shown in table 1. The table shows that $z_s \leq 1$, which lies in the proposed range in the references \cite{hamity1}

\section{Energy Conditions}

In this section we are going to verify the energy
conditions namely null energy condition (NEC), weak energy
condition(WEC), strong energy condition(SEC), at all points in the interior of a star
which will be satisfied if the following inequalities hold simultaneously:
\begin{equation}
NEC: \rho(r)\geq 0 ,
\end{equation}
\begin{equation}
WEC: \rho(r)-p_r(r) \geq  0~~ and
 ~~\rho(r)-p_t(r) \geq  0,
\end{equation}
\begin{equation}
SEC
: ~~\rho-p_r(r)-2p_t(r) \geq  0,
\end{equation}

\begin{figure}[htbp]
    \centering
        \includegraphics[scale=.8]{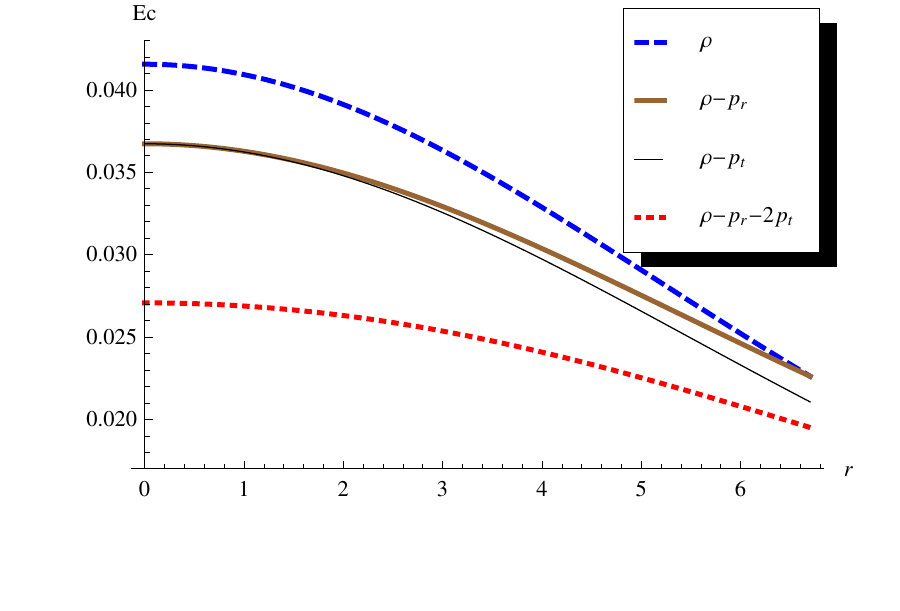}
       \caption{The NEC, WEC and SEC are plotted against r for the compact star Her X-1 by taking the same values of the constant mentioned in fig. 1}\label{ec}
\end{figure}
We will check the energy conditions with the help of graphical representation.
In Fig.\ref{ec}, we have plotted the L.H.S of the above inequalities which
verifies that  all the energy conditions are satisfied
at the stellar interior.

\begin{table*}
\small
\caption{Calculated values of the parameter from the model\label{tbl-2} }
\begin{center}
\begin{tabular}{@{}crrrrrrrrrrr@{}}
\hline
Compact Star  & Mass $(M_{\odot})$   & Radius  & B & C & F \\
              &            &  (Km)       &  & $(km^{-2})$  &$(km^{2})$    \\
\hline
SAX J 1808.4-3658(SS1) &1.435&7.07&0.176 &$4.59\times10^{-3}$&83.48\\
Her X - 1&0.98&6.7&0.382 &$2.34\times10^{-3}$  &104.03\\
Vela~X-12 & 1.77& 9.99& 0.265 &$1.59\times10^{-3}$ &190.94                                                                                                      \\
PSR~ J1614-2230 &1.97&10.3&0.215&$1.82\times10^{-3}$&188.03\\
Cen X - 3&1.49&9.51&0.341&$1.33\times10^{-3}$&195.67\\
\hline
\end{tabular}
\end{center}
\end{table*}

\begin{table*}
\small
\caption{Calculated values of the parameter from the model\label{tbl-2} }
\begin{center}
\begin{tabular}{@{}crrrrrrrrrrr@{}}
\hline
Compact Star  & central density   & central pressure   & compactness & surface redshift\\
              &        $gm/cm^3$    &  $dyne/cm^{2}$       &  &   \\
\hline
SAX J 1808.4-3658 (SS1)& $3.184\times10^{15}$ &$8.192\times10^{35}$&0.2994 &0.5787\\
Her X - 1&$2.232\times10^{15}$&$2.334\times10^{35}$&0.2157    &0.3263\\
Vela~X-12 & $1.317\times10^{15}$&$2.194\times10^{35}$& 0.2613&0.4474                                                                                                     \\
PSR~ J1614-2230 &$1.382\times10^{15}$&$2.894\times10^{35}$&0.2821&0.5148\\
Cen X - 3&$1.220\times10^{15}$&$1.487\times10^{35}$&0.2311&0.3636\\
\hline
\end{tabular}
\end{center}
\end{table*}

\section{Stability of the model}
\subsection{stability under three different forces}
Now we want to examine whether our present model is stable under three forces $viz$ gravitational force, hydrostatics force and anisotropic force which can be described by the following equation
\begin{equation}
-\frac{M_G(r)(\rho+p_r)}{r}e^{\frac{\nu-\mu}{2}}-\frac{dp_r}{dr}+\frac{2}{r}(p_t-p_r)=0,
\end{equation}
 proposed by Tolman–Oppenheimer–Volkov and named as TOV equation.\\
 where $M_G(r) $ represents the gravitational mass within the radius $r$, which can derived from the Tolman-Whittaker
formula and the Einstein's field equations and is defined by
\begin{equation}
M_G(r)=\frac{1}{2}re^{\frac{\mu-\nu}{2}\nu'}.
\end{equation}
Plugging the value of $M_G(r)$ in equation $(32)$, we get
\begin{equation}\label{tov1}
-\frac{\nu'}{2}(\rho+p_r)-\frac{dp_r}{dr}+\frac{2}{r}(p_t-p_r)=0.
\end{equation}
The above expression may also be written as
\begin{equation}
F_g+F_h+F_a=0,
\end{equation}
\begin{figure}[htbp]
    \centering
        \includegraphics[scale=.8]{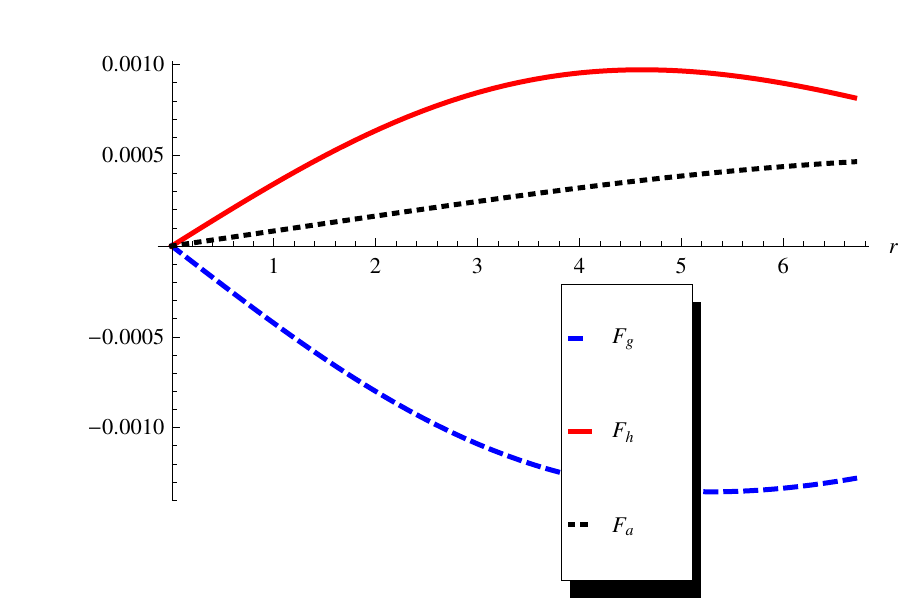}
       \caption{Counter-balancing of three forces acting on the system are shown in the figure for the compact star Her x-1 by taking the same values of the constant mentioned in fig. 1 }\label{tov}
\end{figure}
where $F_g, F_h$ and $F_a$ represents the gravitational, hydrostatics and anisotropic forces respectively.
Using the Eqs. (11-13), the expression for $F_g,F_h$ and $F_a$ can be written as,

\begin{eqnarray}
F_g&=&-\frac{\nu'}{2}(\rho+p_r)=-\frac{32 C^2 r}{8\pi(1 + C r^2)^2 [1 + 64 B C^2 F r^2 (1 + C r^2)^2]^2}f_3(r)\\
F_h&=&-\frac{dp_r}{dr}=\frac{16 C^2 r }{8\pi(1 + C r^2)^2 [1 + 64 B C^2 F r^2 (1 + C r^2)^2]^2}f_4(r)\\
F_a&=&\frac{2 C^2 r [1 - 48 B C F (1 + C r^2)^3 + 512 B^2 C^2 F^2 (1 + C r^2)^6]}{\pi(1 + C r^2)^2 [1 + 64 B C^2 F r^2 (1 + C r^2)^2]^2}
\end{eqnarray}
where,
\begin{eqnarray*}
f_3(r)&=&1 + 16 B C F (1 + C r^2)^2 (1 + 7 C r^2)\\
f_4(r)&=&1 - 512 B^2 C^2 F^2 (1 + C r^2)^6 +
   16 B C F (1 + C r^2)^2 (5 + 17 C r^2)
\end{eqnarray*}

The profile of three different forces are plotted in fig. \ref{tov}. The figure shows that gravitational force is dominating is nature and is counterbalanced by the combine effect of hydrostatics and anisotropic force.

\subsection{Sound Velocity}
In this section we are going to find the subliminal velocity of sound. For a physically acceptable model of anisotropic fluid sphere the radial and transverse velocity of sound should be less than 1 which is known as causality conditions.\\
The radial velocity $(v_{sr}^{2})$ and transverse velocity $(v_{st}^{2})$ of sound can be obtained as:
\begin{eqnarray}
v_{sr}^{2}&=&\frac{dp_{r}}{d\rho}=\frac{1 + 64 B C^2 F r^2 (1 + C r^2)^2}{16 B C F (1 + C r^2)^2 f_6(r)}f_5(r)\\
v_{st}^{2}&=&\frac{dp_{t}}{d\rho}=\frac{r^2 + 8 B F (1 + C r^2)^2 }{8 B F (1 + C r^2)^3 f_6(r)}f_7(r)
\end{eqnarray}
where,
\begin{eqnarray*}
f_5(r)&=&1 +
   16 B C F (1 + C r^2)^2 \{5+17 C r^2 -32 B C F (1 + C r^2)^4\}\\
f_6(r)&=&2048 B^2 C^3 F^2 r^2 (1 + C r^2)^6 -(5+7 C r^2)+\\
   &&32 B C F (1 + C r^2)^2 \{5 + C r^2 (16 + 23 C r^2)\}\\
f_7(r)&=&8 + C r^2 (39 + 55 C r^2) -
    64 B C F (1 + C r^2)^3 \{1+2 C r^2 (1-C r^2)\}
\end{eqnarray*}

 \begin{figure}[h!]
    \centering
        \includegraphics[scale=.55]{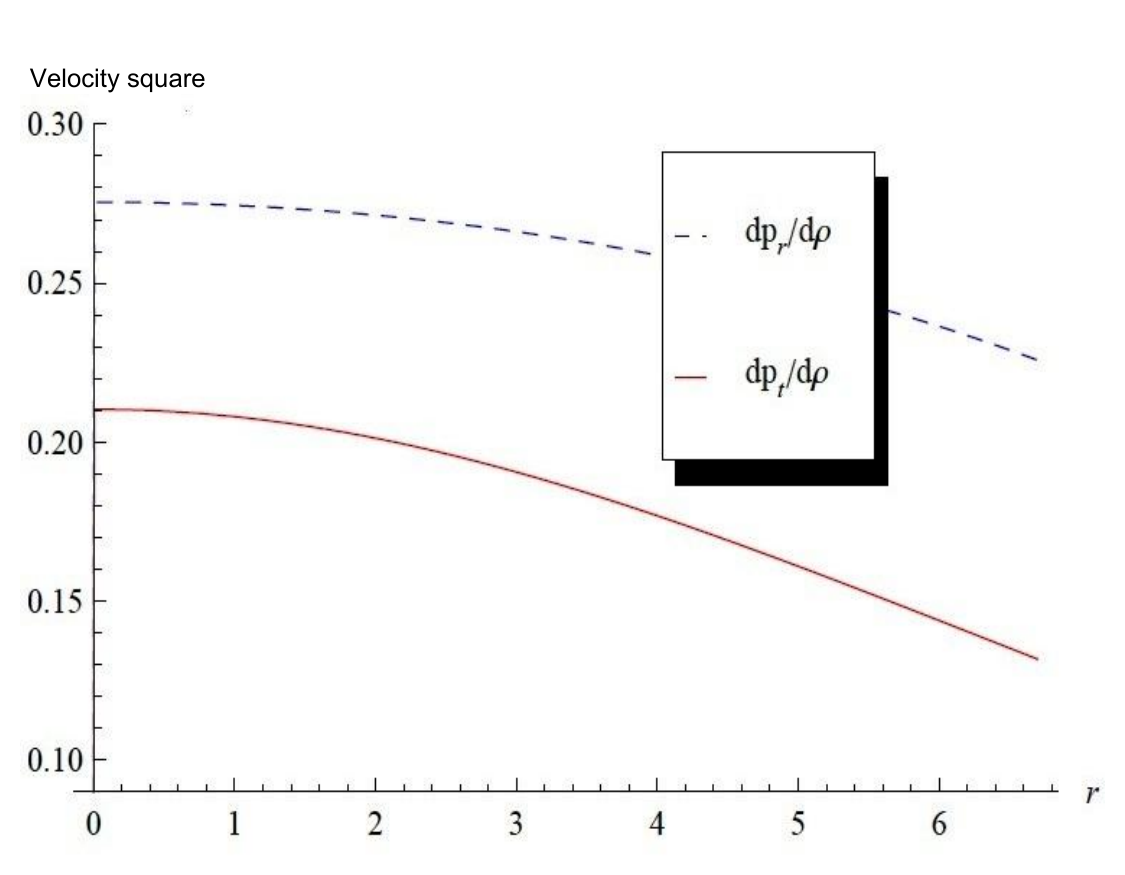}
       \caption{square of radial and transverse velocity of sound are plotted against r for the star Her X-1 by employing the same values of the arbitrary constants as mentioned in fig. 1}\label{sv1}
    \label{fig:12}
\end{figure}

  \noindent The profile of radial and transverse velocity of sound have been plotted in fig. \ref{sv1}, the figure indicates that our model satisfies the causality condition. In 1992 Herrera \cite{her} proposed the `cracking' method to study the stability of anisotropic stars under the radial perturbations. Using the concept of cracking Abreu {\em et al.}\cite{ab} proved that the region of an anisotropic fluid sphere where $-1\leq v_{st}^{2}-v_{sr}^{2}\leq0$ is potentially stable but the region is potentially unstable where $0\leq v_{st}^{2}-v_{sr}^{2}\leq1$. Moreover for an anisotropic model of compact star $|v_{st}^{2}-v_{sr}^{2}|<1$ according to Andr\'{e}asson \cite{andre}. Fig.\ref{sv2} and Fig. \ref{sv3} indicates that cracking method and Andr\'{e}asson's condition are verified.

\begin{figure}[h!]
    \centering
        \includegraphics[scale=.7]{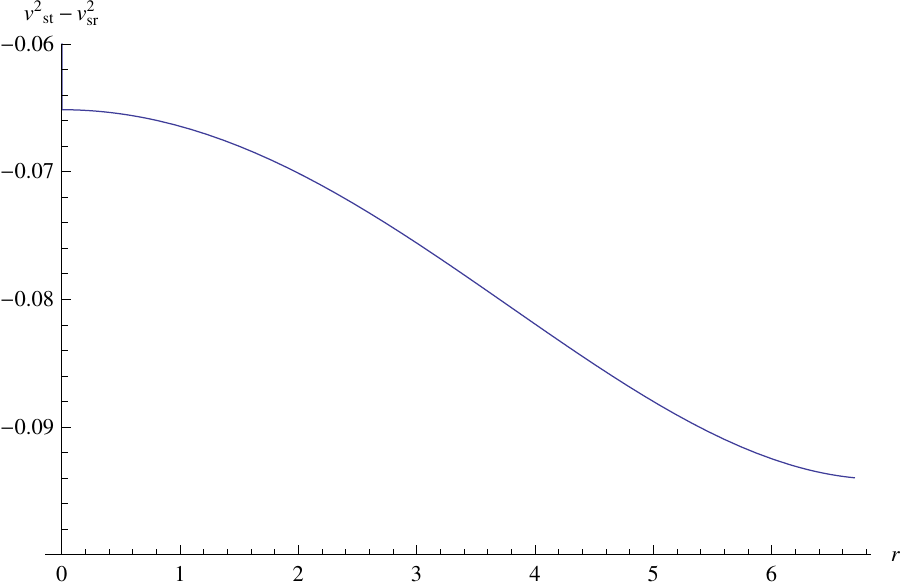}
       \caption{$v_{st}^{2}-v_{sr}^{2}$ are plotted against r for the star Her X-1 by employing the same values of the arbitrary constants as mentioned in fig. 1}\label{sv2}
\end{figure}

\begin{figure}[h!]
    \centering
        \includegraphics[scale=.7]{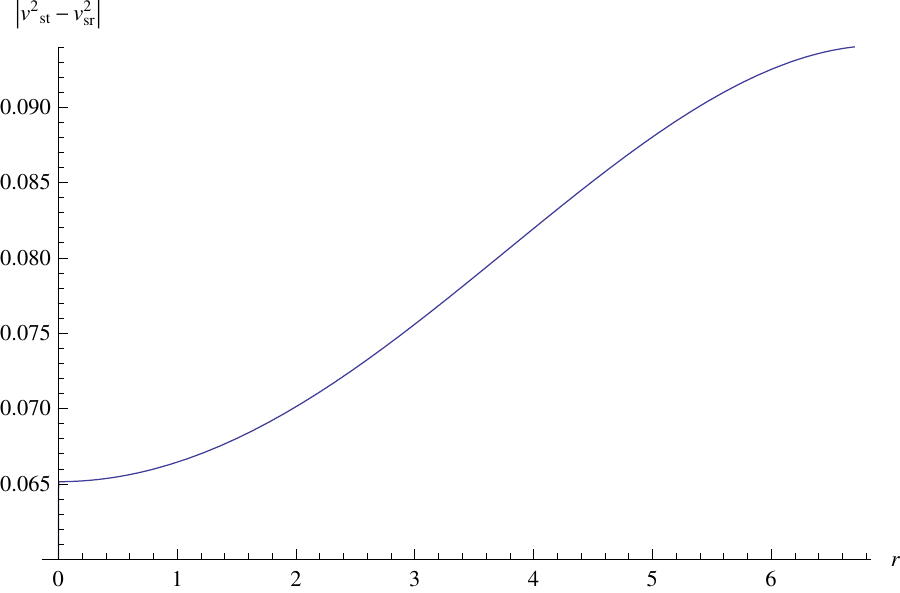}
       \caption{$|v_{st}^{2}-v_{sr}^{2}|$ are plotted against r for the star Her X-1 by employing the same values of the arbitrary constants as mentioned in fig. 1}\label{sv3}
\end{figure}

\subsection{Adiabatic index}
It is well known, the condition for the stability of a Newtonian isotropic sphere is $ \gamma>4/3$ (see \cite{bondi} for a detailed discussion on this point)
This condition changes for a relativistic isotropic  sphere \cite{bondi}, and more so for an anisotropic general relativistic sphere \cite{1,2}
Where $\Gamma$ is called the adiabatic index and is given by,
\begin{eqnarray}
  \Gamma_r &=& \frac{\rho+p_r}{p_r}\frac{dp_r}{d\rho}   \nonumber  \\
  &=& \frac{[1 + 16 B C F (1 + C r^2)^2 (1 + 7 C r^2)] [1-
   16 B C F (1 + C r^2)^2 f_8(r)]}{16 B C F (1 + C r^2)^2 [1-8 B C F (1 + C r^2)^3]G(r)}\\
  \Gamma_t &=& \frac{\rho+p_t}{p_t}\frac{dp_t}{d\rho}   \nonumber  \\
  &=& \frac{[r^2 + 8 B F (1 + C r^2)^2 f_9(r)]H(r)}{8 B F (1 + C r^2)^3 \{1 + 2 C r^2 -
    8 B C F (1 - C r^2) (1 + C r^2)^3\}G(r)}
\end{eqnarray}

\begin{eqnarray*}
f_8(r)&=&\{5+17 C r^2- 32 B C F (1 + C r^2)^4\}\\
f_9(r)&=&\{8 + C r^2 (39 + 55 C r^2) -
      64 B C F (1 + C r^2)^3 (1 + 2 C r^2 (1 - C r^2))\}\\
G(r)&=&
   2048 B^2 C^3 F^2 r^2 (1 + C r^2)^6 +
   32 B C F (1 + C r^2)^2 [5 + C r^2 (16 + 23 C r^2)]\\
   &&-(5+7 C r^2)\\
H(r)&=&1 + 2 C [r^2 +
      8 B F (1 + C r^2)^3 \{1 + 4 C r^2 (1 + 8 B C F (1 + C r^2)^3)\}]
\end{eqnarray*}
For a relativistic anisotropic fluid sphere the stability condition is given by,
\begin{equation}
\Gamma_r,\Gamma_t>\frac{4}{3}+\left[\frac{4}{3}\frac{(p_{t0}-p_{r0})}{|p_{r0}^\prime|r}+4\pi\frac{\rho_0p_{r0}}{|p_{r0}^\prime|}r\right],
\end{equation}
where, $p_{r0}$, $p_{t0}$, and $\rho_0$ are the initial radial, tangential, and energy density in static equilibrium satisfying eq. (\ref{tov1}). The first and last term inside the square brackets, the anisotropic and relativistic corrections respectively, being positive quantities, increase the unstable range of adiabatic index. To see the behavior of the adiabatic index we have plotted $\Gamma_r$ and $\Gamma_t$  vs. r in fig \ref{g1} and \ref{g2} respectively. The figures show that both $\Gamma_r,\Gamma_t>\frac{4}{3}$ everywhere within the stellar interior.
\begin{figure}[htbp]
    \centering
        \includegraphics[scale=.7]{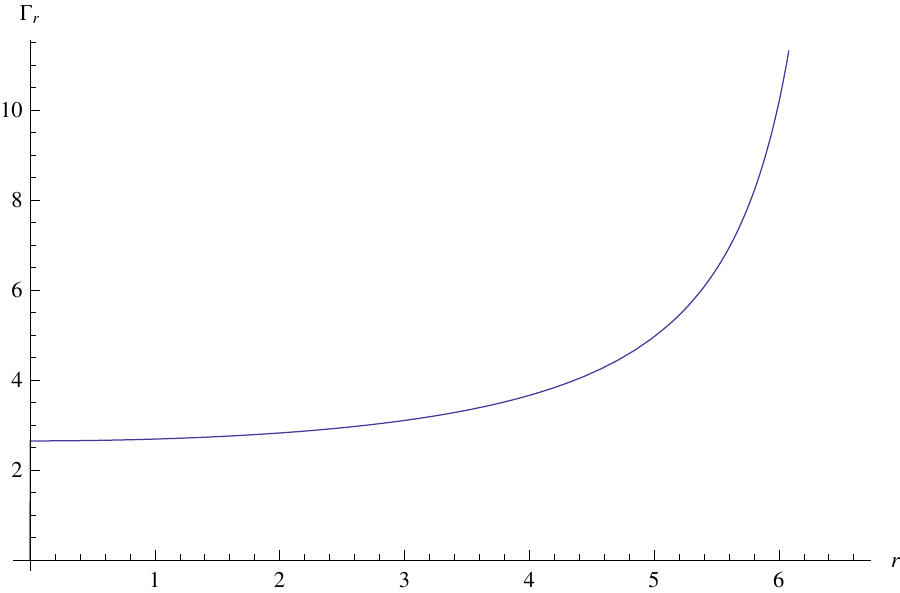}
       \caption{$\Gamma_r$ is plotted against r inside the stellar interior for the compact star Her X-1 by employing the same values of the arbitrary constants as mentioned in fig. 1}\label{g1}
\end{figure}
\begin{figure}[htbp]
    \centering
        \includegraphics[scale=.7]{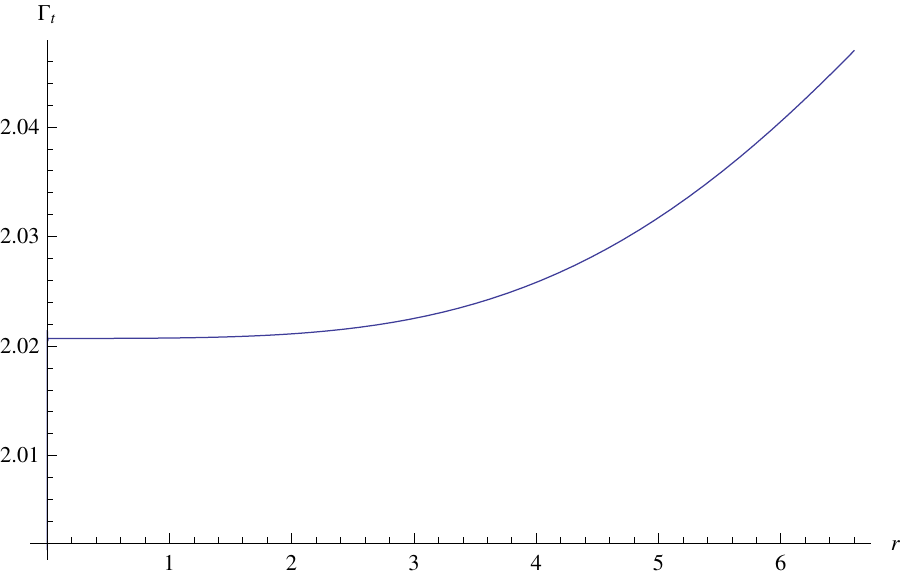}
       \caption{$\Gamma_t$ is plotted against r inside the stellar interior for the compact star Her X-1 by employing the same values of the arbitrary constants as mentioned in fig. 1}\label{g2}
\end{figure}

\section{Discussion and conclusion}
In the problem of obtaining anisotropic fluid  at least two conditions must be provided from outside one is in fact anisotropic factor measuring the anisotropy of the system or one of the metric potential and the other rests on our choice. In the present case we have considered class one conditions (Karmarkar conditions) and metric potential $g_{44}=e^{\nu}$. It is worth pointing out here that the possibility of perfect fluid distribution is ruled out as only solutions satisfying Karmarkar conditions are either Schwarzschild's interior solution or Kohlar Chao solution. So for this purpose, We have started the metric function $e^{\nu}$ which is entirely different from Schwarzschild's interior solution or Kohlar Chao solution. After that we have obtained the metric potential $e^{\lambda}$ by using the class 1 condition(Eq.4), which is
regular and monotone increasing away from centre of the star. \\
The physical features of the anisotropic star as follows:\\

1.The density $\rho$ and radial pressure $p_r$  and tangential
pressure $p_t$ are positive and monotonically decreasing
functions of radial coordinate $r$ for $0\leq r\leq r_b$ where
$r_b$ is the radius at the boundary of the star.
Also radial pressure vanishes at the surface of the star as  expected.\\
2. The derivatives of radial pressure and density are negative
confirming the fact that radial pressure and density are monotone
decreasing functions of $r$.\\
3. The anisotropic factor  $\Delta$ is positive and monotone
increasing throughout the stellar interior,implying $p_t>p_r$ for
$0\leq r\leq r_b$. However $\Delta=0$ at the centre of the star.
The anisotropic force is repulsive in
nature.\\
4. The calculated expression of stellar mass is regular, positive
and monotone increasing function of $r$ for $0\leq r\leq r_b$ and
it vanishes at the stellar centre.\\
5. It can be seen from fig. \ref{ur} that the compactness factor
satisfies Buchdahl\cite{buch} inequality, i.e,
$\frac{2M}{r_b}<\frac{8}{9}$.
Also it is clear from Table 2 that the value of surface redshift function lies in the range $z_s<1$ .\\
6. The three main energy conditions ,viz.,null energy condition(NEC), weak energy condition(WEC) and the strong energy condition(SEC) are all satisfied by our star model.\\
7. Fig. \ref{tov} clearly shows that the system is in  static
equilibrium under the action of hydrostatic,anisotropic and
gravitational forces. Here the gravitational force is dominating
is nature and is counterbalanced by the combined
effect of hydrostatics and anisotropic forces.\\
8. The radial and transverve subliminal velocity of  sound for
our model satisfies the causality condition, i.e.,
$v_{sr},v_{st}<1$ which acertains that our model is stable.
Also since $|v_{sr}^{2}-v_{st}^{2}|<1$ everywhere inside the anisotropic fluid sphere, our model is also stable according to Andr\'{e}asson\cite{andre}.\\
9. From the plot of  adiabatic index for our model we see that both $\Gamma_r$ and $\Gamma_t$ $>\frac{4}{3}$ everywhere within the interior of the fluid sphere and hence our model is stable.\\
From table 1 we claim that this model is compatible with the
compact objects Her X-1, SAX J 1808.4-3658(SS1), Vela X-12, PSR
J1614-2230 and Cen X-3 and it is free from cental singularity.
Hence we can claim that the physical features of our model are
quite reliable and physically feasible.

\section*{Acknowledgments}
SKM acknowledges support this research work from the authority of University of
Nizwa, Nizwa, Sultanate of Oman. The authors are also thankful to anonymous referees for
raising several pertinent issues which have helped us to improve
the manuscript substantially.

\end{document}